\documentstyle[12pt]{article}
\font\big=cmr12
\def\beqn{\begin{equation}}
\def\eeqn{\end{equation}}
\def\beqna{\begin{eqnarray}}
\def\eeqna{\end{eqnarray}}
\def\T{k_B T}
\def\x{{\vec x}}

\def\f{{\vec f}}

\def\k{{\vec k}}
\def\A{{\vec A}}
\def\B{{\vec B}}
\def\E{{\vec E}}
\def\j{{\vec j}}
\def\H{{\vec H}}
\def\r{{\vec r}}

\def\x{\hbox{\bf x}}
\def\u{\hbox{\bf u}}

\def\eps{\epsilon}

\def\pt{{\partial}}

\def\o{\over}
\begin{document}
\begin{center}
\big
{\bf Statistical Mechanics of Vortices in Type-II Superconductors}\\
\vskip 2cm
{David A. Huse$^{(1)}$ and Leo R. Radzihovsky$^{(2)}$}\\
\vskip 1cm
{$^{(1)}$ AT\&T Bell Laboratories, Murray Hill, New Jersey 07974}\\
{$^{(2)}$ The James Franck Institute, Physics Department}\\
{University of Chicago, 5640 South Ellis Avenue, Chicago, IL 60637}\\
\end{center}
\vskip 4cm
\begin{abstract}
Thermal fluctuations and disorder play an essential role in high-T$_c$
superconductors. After
reviewing the mean-field phase diagram we describe significant
modifications that result when the effects of finite temperature
and disorder are incorporated.
Thermal fluctuations cause the
melting of the Abrikosov flux line lattice into a vortex liquid
at high temperatures, and disorder produces novel glass phases at low
temperatures.
We analyze phase transitions between these new
phases and describe transport properties in various regimes.
\end{abstract}
\newpage

\section{INTRODUCTION}

The microscopic mechanism of superconductivity in the new high-T$_c$
materials has remained a mystery despite much vigorous recent
activity\ \cite{Phillips89}. However significant
progress in our understanding of the behavior of these materials has
been achieved through phenomenological approaches. The phenomenological
theory of superconductors was introduced in the
1950's by Landau and has
been successful in describing the conventional superconductors\ \cite{Parks69}.
Many of the key properties and phenomena in high-T$_c$ materials can also be
described entirely with the conventional Ginzburg-Landau theory, albeit with
unconventional values of parameters. In this new range of parameters
an understanding of the interplay of strong thermal fluctuations and disorder,
both in statics and dynamics, is required.

Within a mean-field treatment, appropriate to
conventional (low-T$_c$) type II superconductors, the Ginzburg-Landau theory
leads to the celebrated Abrikosov vortex lattice\ \cite{Abrikosov57} and the
Meissner superconducting phases, with well-defined continuous transitions
occurring at the upper and lower critical fields $H_{c2}(T)$ and $H_{c1}(T)$,
respectively (see Fig.\ref{fig1}(a)). However, unlike
these conventional type II materials, where in the
absence of disorder the flux lines always arrange themselves into an ordered,
static triangular lattice, high-T$_c$ superconductors exhibit much richer
vortex-line states. Because of the combination of high T$_c$ ($\sim 100$ K),
short coherence length $\xi$, large penetration length $\lambda$, and large
anisotropy, the Abrikosov lattice of flux lines is destabilized at finite
temperatures by the thermal fluctuations and disorder.

The resulting magnetic field versus temperature (HT) phase diagram for
high-T$_c$ materials is drastically
modified by the fluctuations and disorder. In Fig.\ref{fig1}(b) we have
schematically illustrated a resulting phase diagram for the high-T$_c$
materials, with various new novel phases and regimes that have been
proposed.  The fluctuation and disorder-corrected phase diagram is
remarkably rich and is to be contrasted with mean-field phase diagram
characterizing the conventional superconductors, displayed in
Fig.\ref{fig1}(a).

At low temperatures, the translational order of the
Abrikosov lattice is lost due to disorder that
appears in the form of oxygen vacancies and interstitials, and grain and twin
boundaries\ \cite{Larkin79}. Vortex glass\ \cite{FFH89}, Bose
glass\ \cite{NelsonVinokur92} and polymer glass\ \cite{Obukhov90}
are some of the new low-temperature phases that have been proposed to replace
the traditional Abrikosov flux lattice. Even in pure crystals
thermal fluctuations have been proposed to lead to a novel supersolid vortex
phase\ \cite{GlazmannKoshelew90,Frey93}, which
unlike the 3d Abrikosov lattice, is a quasi-2d lattice with freely wandering
interstitials. At even higher temperatures or higher
fields the flux lattice melts and is replaced by a vortex liquid of highly
mobile and flexible vortex lines\ \cite{NelsonSeung89,Gammel88}. Motivated by
recent experiments, in which a resistive ``shoulder'' was observed, proposals
have been made that the vortex liquid phase could possibly be further
subdivided into distinct liquid phases, and can undergo transitions between
hexatic\ \cite{Marchetti90,Worthington90,NelsonHalperin79} and
isotropic, and entangled and disentangled vortex liquids\ \cite{NelsonSeung89}.

To study the transport properties in these new novel phases the
interaction between the transport current, the screening supercurrents and
the magnetic field must be
understood. These interactions lead to a Magnus-Lorentz force
exerted on the vortex line perpendicular to the transport current.
A motion of vortices in response to this force
generates finite voltages and resistivity,
and therefore leads to a breakdown of true superconductivity.

The Meissner phase,
in which the flux is completely expelled, is fully superconducting up to a
critical surface current density $j_c$ beyond which thermal fluctuations
can produce
vortex loops, which expand under the influence of the transport current.
In the ideal situation an unpinned vortex lattice can move
as a whole in response to a uniform current and therefore is not
truly superconducting. However, the
vortex lattice has no resistive linear response for periodic current patterns
with a finite wavelength because of the finite shear modulus of the lattice.
Therefore the vortex lattice exhibits a novel fully nonlocal
linear resistivity. Of course in practice the vortex lattice will have a
superconducting linear response even to a uniform current, since the sample
boundaries and even weak disorder will pin the lattice in place.
Newly proposed vortex glass phases are also believed to be fully
superconducting. The vanishing of linear resistivity in the Meissner and
the putative glassy solid phases (vortex and Bose glass) is related to the
divergence of energy barriers (with vanishing current) separating the ground
state and the low-lying excited states.

In the vortex liquid phase the flux lines can more easily respond to the
transport current-induced forces, although interactions with other lines
are believed to lead to a much higher viscosity than that of the conventional
isotropic liquids\ \cite{NelsonSeung89,Obukhov90}. A phenomenological
 hydrodynamic theory can be used to
describe the resulting flux-flow resistivity which is nonlocal with
properties intermediate between that of the fully superconducting vortex
lattice and the normal phase\ \cite{Marchetti91,Radzihovsky93}.

The mean-field phase transitions are also modified by strong thermal
fluctuations. In contrast to low-T$_c$ materials, in the $Cu O_2$ based
superconductors there is no well defined thermodynamic transition
at the upper critical field, $H_{c2}$.  The mean-field transition is
replaced by a smooth crossover (indicated by a dashed line in
Fig.\ref{fig1}(b))
from the normal state to the non-superconducting, but more
conductive vortex liquid regime.
Therefore the distinction between the vortex liquid and the
normal state is only quantitative.
As $H^{MF}_{c2}(T)$ is passed from above the
resistivity decreases, the vortices begin to form as the
fluctuations in the superconducting order parameter grow to a nonlinear
regime.  However, because of the motion of vortex lines, which destroys
phase coherence, the average value of the superconducting order parameter
$\psi(\r)$ is still zero in vortex liquid regime as it is in the
normal state.

The real transition occurs only at lower fields and temperatures,
where superconducting vortex solid phases form. For clean samples a
first order freezing transition transforms the liquid into the
vortex lattice. Despite the lack of a full theory of this 3d melting
transition the location of the melting curve $B_m$(T) can be estimated by
the Lindemann criterion\ \cite{NelsonSeung89,Houghton89,Brandt89}. However, in
the presence of disorder and strong vortex
line interactions physical arguments have predicted a transition to the
highly correlated fully superconducting isotropic and anisotropic
(Bose glass) vortex glass phases. The vortex glass transition is believed to
replace the melting transition of pure samples.  Although no fundamental
analytical description is available to date, this continuous glass
transition can be analyzed using scaling theory that is in good
agreement with transport experiments.

Recently, there have been several experimental observations of the transition
into a glassy vortex phase that agree with predictions of the scaling theory
\ \cite{Koch89,Gammel91,Sandvold92,Dekker92}. The impressive data collapse
over several decades, from which the critical exponents characterizing
the transition can be determined, provides good evidence for the existence
of the vortex glass phase. This data and analytical analysis is further
supported by various numerical studies\ \cite{Reger91,Cieplak,Gingras92}.

In these lectures we will review some of the highlights of the phenomenology of
high-T$_c$ superconductors. We begin in Sec. 2 by reviewing the picture that
emerges from the mean-field analysis, describing the equilibrium and
non-equilibrium properties of the resulting phases and the mean-field
transitions.  As we described above, effects of thermal fluctuations and
disorder are very important and in Sections 3 and 4 we describe how these
effects modify the HT phase diagram. After describing equilibrium properties
of the new vortex phases, we proceed to analyze their transport properties.
We apply a scaling theory to predict the form of the resistivity in the
superconducting phases. In Sec. 5 we describe the nature of the
fluctuation-corrected phase transition and again use scaling theory
to summarize its properties in terms of a small set of universal
critical exponents, which can be determined through numerical and
experimental means. In each section we compare theoretical results
with experiments and numerical studies, and for the most part find a
coherent and consistent picture emerging.

\section{MEAN-FIELD THEORY}

Irrespective of the microscopic mechanism responsible for the attractive
interaction that leads to the binding of the electrons into Cooper pairs,
the Ginzburg-Landau theory postulates the existence of an
order parameter, a complex scalar field $\psi(\vec r)$ that is the condensate
``wavefunction'' of the bosonic Cooper pairs of electric charge $2e$.
The long wavelength properties of the superconductors are then
encoded in the Ginzburg-Landau free energy functional that is
assumed to have a local expansion in powers of $\psi(\vec r)$ and
its spatial gradients. Near the superconducting transition, $T_c$,
the order parameter $\psi(\vec r)$ is small, and it is sufficient to
retain only the lowest order terms in the expansion of the free
energy\ \cite{HigherOrder}
\beqna\label{eq:FreeEnergy}
&F_{GL}=\int d^3 r&\left[{\hbar^2\over 2m_z}\left|\left({\pt\o\pt z}-
{i2e\over\hbar c}A_z(\r)\right)\psi(\r)\right|^2\right. \nonumber \\
&&\left.+{\hbar^2\over 2m}\left|\left(\vec\nabla_\perp-
{i2e\over\hbar c}\A_\perp(\r)\right)\psi(\r)\right |^2\right. \nonumber \\
&&\left. + \alpha|\psi(\r)|^2+{1\o 2}\beta|\psi(\r)|^4\right. \nonumber \\
&&\left. + {1\over 8\pi}\left(\H-\vec\nabla\times\A(\r)\right)^2\right]\;.
\eeqna
The first two terms are the supercurrent kinetic energy.  The sum of the third
and fourth terms is the local pairing energy, approximated at quartic order.
The last term is the magnetic field energy that couples the
external magnetic field to
the magnetic induction $\B(\r)=\vec\nabla\times\A(\r)$.
Physically the parameters $m_z$ and $m$ in the above expansion
can be identified with the effective mass of the Cooper pair along the
$c$-axis (chosen in $z$-direction) and in the $ab$-plane. With the
anisotropy parameter $\gamma\equiv \sqrt{m/m_z}\ll 1$
we can describe the large material anisotropy of the high-T$_c$
layered materials, which are much stronger superconductors in the $ab$-plane
than along the $c$-axis\ \cite{Kogan81}. The
$\alpha$ and $\beta$ are the material parameters that depend
on temperature $T$. Generally $\alpha(T)\approx a(T-T_c^{MF})/T\equiv t$
while the other parameters have only a smooth dependence on $T$  without
sign changes.  This dependence
and the general validity of above description has been verified by
Gorkov\ \cite{Gorkov} through his derivation of the Ginzburg-Landau
theory from the microscopic Bardeen-Cooper-Schreiffer (BCS) theory of
conventional superconductors\ \cite{Parks69}.

The mean-field description which neglects the effects
of fluctuations is valid only outside the critical region around the
normal-superconducting (NS) transition. For conventional,
low-temperature  bulk superconductors in zero magnetic field, this critical
region is unmeasurably small with fluctuations only becoming
important for $(T-T_c)/T_c < 10^{-7}$. On the other hand the
combination of high-T$_c$, strong anisotropy, short coherence
length $\xi$ and large magnetic penetration length $\lambda$ of
high-T$_c$ materials expands the critical region by several orders
of magnitude. As we will see, thermal fluctuations will lead to
significant modifications of the mean-field description that we
first present below.

\subsection{Mean-field statics}

The mean-field theory description is obtained by simply minimizing
$F_{GL}[\psi,\A]$ with respect to $\psi(\r)$ and $\A(\r)$.
The resulting mean-field
phase diagram is displayed in Fig.\ \ref{fig1}(a), with three distinct phases.
\begin{figure}
\caption{(a) Mean-field phase diagram of type II superconductors. (b)
Schematic picture of a phase diagram of high-T$_c$ superconductors which
includes effects of thermal fluctuations and disorder.}
\label{fig1}
\end{figure}
For $T>T_c$ i.e. $\alpha(T)>0$ the pairing and magnetic terms are
minimized by $\psi(\r)=0$ and $\B(\r)=\H$ characteristic of the
normal state. In the Meissner phase the
$U(1)$ original symmetry of $F_{GL}$
($\psi(\r)\rightarrow\psi(\r)e^{i\phi}$) is spontaneously
``broken''\ \cite{SymmBrake} by $\psi(\r)=\sqrt\rho_s=\sqrt{|\alpha|/\beta}$
\beqn
\langle\psi^*(\r)\psi(\r+\vec R)\rangle\rightarrow\rho_s\;,\;\;\;
\mbox{for $|\vec R|\rightarrow\infty$}\;,
\eeqn
which leads to ``off-diagonal'' long-range order that sets in
exponentially over the coherence length $\xi=\hbar/\sqrt{2m|\alpha|}$
in the $ab$-plane. Along the $z$-axis the coherence length is much
shorter, $\xi_z=\gamma\xi$, scaled down by the anistropy parameter,
which can be as small as $1/100$ in $BSCCO$.

Borrowing the notation from superfluids we associate the square of
the condensate amplitude with the density of the superconducting
pairs. It plays the role of stiffness
for the phase $\phi(\r)$ variations, as we will see below.
The coherence length over which $\psi(\r)$ varies (due to for example
imposed boundary conditions) is set by the competition
between the kinetic energy and the pairing energy and reflects the
size of the Cooper pair.

The magnetic field $\B(\r)=0$ is
expelled everywhere in the sample except for the boundary where
it leaks in a distance $\lambda=\sqrt{m c^2\beta/16\pi|\alpha|e^2}$ along the
$ab$-plane. The superconducting screening currents along the $z$-direction
are less effective in keeping the applied field from penetrating along the
$c$-axis, with $\lambda_z=\lambda/\gamma$. The London penetration length
$\lambda$ is set by the competition between the magnetic field term and the
kinetic energy due to the screening currents.

For vanishing applied field the transition is in the same
universality class as that of paramagnetic-ferromagnetic transition
described by the XY model, with disordered paramagnetic state and
spin aligned correlated state corresponding to the normal and
Meissner states, respectively.

As was predicted by Abrikosov\ \cite{Abrikosov57}, for type II
superconductors, upon
increasing the applied field beyond the lower critical field $H_{c1}$
a type II superconductor undergoes a transition from the Meissner phase to
the Abrikosov vortex lattice phase. In the vortex lattice phase
the magnetic field penetrates  the material
and induces elementary topological defects in the order parameter $\psi(\r)$,
which in three dimensions are vortex lines.
In the Abrikosov phase the long-range order persists with
$\psi(\r)=|\psi(\r)|e^{i\phi(\r)}$. However, instead of the
phase being constant everywhere, it ``winds'' by $2\pi$
around a vortex line.  The amplitude of $\psi(\r)$
is nearly constant everywhere away from the vortex core but is
substantially suppressed inside the core of size $\xi$ or $\xi_z$, for fields
applied perpendicular to the $ab$-plane or $z$-axis, respectively. The
magnetic flux per vortex is the flux quantum,
$\phi_0=h c/2 e$. When the vortices are well separated this flux is confined
by circulating
screening currents within a London penetration length $\lambda$ of the core.
For an isolated straight flux line running parallel to an axis of cylindrical
symmetry of the material the field is of the form.

\beqn\label{eq:Bggll}
\B(\r)\approx\left\{
\begin{array}{ll}
e^{-r/\lambda} & \mbox{for $r\gg\lambda$}\\
-\ln(r/\lambda)  & \mbox{for $\xi\ll r\ll\lambda$}\;\;\;.
\end{array}\right.
\eeqn
For the most common situation of magnetic field applied along the $c$-axis
a vortex line has structure on two length scales, $\xi$ and
$\lambda$, and its crossection in $ab$-plane is
illustrated in Fig.\ \ref{fig2}
\begin{figure}
\caption{Crossection of a vortex line in type II superconductors,
illustrating the relevant length scales discussed in the text.}
\label{fig2}
\end{figure}

In the Meissner phase and in the Abrikosov phase far away from the core
the amplitude of the order parameter is a constant
$|\psi(\r)|=\sqrt{\rho_s}$ that minimizes the pairing energy.
Only the phase $\phi(\r)$ varies substantially. Using this
{\it Ansatz} for the order parameter $\psi(\r)=\sqrt{\rho_s} e^{i\phi(\r)}$
inside the Ginzburg-Landau free energy we obtain the London free energy
which governs the phase and magnetic field variations,
\beqna\label{eq:LondonF}
&F_L=
\int d^3 r&\left[{\hat{\rho}^s_z\o 2}\left({\pt\phi(\r)\o\pt z}-
{2\pi\o\phi_0}\A_z(\r)\right)^2\right.\nonumber\\
&&\left. + {\hat{\rho}^s\o 2}\left(\vec\nabla_\perp\phi(\r)-
{2\pi\o\phi_0}\A(\r)\right)^2\right.\nonumber\\
&&\left. + {1\over 8\pi}\left(\H-
\vec\nabla\times\A(\r)\right)^2\right]\;,
\eeqna
where $\rho^s$ and $\rho^s_z$ are the normalized eigenvalues of the superfluid
density tensor.
\beqn
\hat{\rho}^s_i=
\left({\hbar^2\o 2m_i}\right)\rho_s={\phi_0^2\o 16\pi^3\lambda_i^2}\;,
\eeqn
where index $i$ labels $x,y,z$ and $\lambda_x=\lambda_y=\lambda$.

The ratio $\kappa=\lambda/\xi$ determines whether
the superconductor is type I (for which the Abrikosov phase
does not exist) or type II according to whether $\kappa < 1/\sqrt2$
or $\kappa > 1/\sqrt2$, respectively. The high-T$_c$ $Cu O_2$
materials are strongly type II, characterized by very large $\kappa$
At low temperatures the typical values of $\xi<15{\rm\AA}$ and
$\lambda >1000{\rm\AA}$ leading to $\kappa > 100$.

The excess Gibb's free energy per unit of length for putting
in $N$ straight, noninteracting vortex/flux lines along the $z$-direction
is given by\ \cite{deGennes}
\beqn\label{eq:nonintGibbs}
\delta F_{line}=N\left[\left({\phi_0\o
4\pi\lambda}\right)^2\ln\kappa-{H\phi_0\o 4\pi}\right]\;,
\eeqn
where $H$ is the $z$-component of applied field $\H$.
This Gibb's energy vanishes at the
lower critical field $H=H_{c1}$ \ \cite{SudboComment},
\beqn\label{eq:Hc1}
H_{c1}={\phi_0\o 4\pi\lambda^2}\ln\kappa\;.
\eeqn
In mean-field theory both the
coherence and the penetration length diverge as $T_c$ is
approached from below, $\xi(T)\sim\lambda(T)\sim(T_c-T)^{-1/2}$.
Since this leads to $\kappa$ that has a smooth behavior near $T_c$
we observe from Eq.\ref{eq:Hc1} that $H_{c1}$ vanishes linearly
as $T$ approaches $T_c$.

For $H_{c1}<H<H_{c2}$ vortices parallel to $\H$
begin to enter the sample with
increasing applied field. The equilibrium
number of lines is determined by the balance between the
noninteracting contribution to Gibb's free energy, Eq.\ref{eq:nonintGibbs},
which is linear in
$N$, and the contribution nonlinear in $N$, coming from the
interaction between flux lines. The vortex-vortex interaction
between two parallel straight lines separated by distance $r>\xi$ is
repulsive and is given by
\beqn\label{eq:interaction}
U(\r)={{\phi_0}^2\over8\pi^2\lambda^2}\left[K_0(r/\lambda)
-K_0(r/\xi)\right]\;,
\eeqn
where $K_0(x)$ is the modified Bessel function with the asymptotics,
\beqn\label{eq:K0asympt}
K_0(x)\approx\left\{
\begin{array}{ll}
({\pi\over2x})^{1/2}e^{-x} & \mbox{for $x\rightarrow\infty$}\\
         -\ln(x)  & \mbox{for $x\rightarrow 0$}\;\;\;.
\end{array}\right.
\eeqn
In Eq.\ref{eq:interaction} we have introduced a short distance
cutoff for $r<\xi$, with the interaction energy saturating at
the value of excess energy for creating a doubly quantized vortex.

In the absence of thermal fluctuations the flux lines crystallize
into a hexagonal Abrikosov vortex lattice, a $3$-d crystal of
infinitely long parallel vortex lines. The vortex lattice breaks
continuous translational symmetry perpendicular to $\H$ and
rotational symmetry about $\H$, as well as the global $U(1)$
symmetry. The flux line density $n$ is directly related to the
average magnetic field $n\approx 1/a^2=\langle\B(\r)\rangle/\phi_0$.
Several values of lattice constants $a$, and the corresponding magnetic
field strengths are displayed in Table~\ref{table1}. For weak
magnetic fields
resulting in flux line separation $a\gg\lambda$, $\xi$ the amplitude
of the order parameter is $\sqrt\rho_s$ and $\B=0$ away from the
vortex cores. However as $H\rightarrow H_{c2}^-$, the amplitude
$|\psi|$ is suppressed while $\B$ grows continuous between vortices,
until at $H_{c2}$ the vortex cores overlap and in the absence of
fluctuations a mean-field transition occurs from the Abrikosov
superconducting phase to the normal phase.

\begin{table}[hbt]
\caption{Relation between typical values of $B$ and lattice
constant $a$}
\vspace{0.5cm}
\label{table1}
\begin{tabular*}{10.7cm}
{l@{\hspace{1.4cm}}l@{\hspace{1.1cm}}l@{\hspace{1.1cm}}l}
\hline
$B$  & $20$ Gauss  & $2$ kGauss & $20$ Tesla\\
$a$  & $1$ $\mu{\rm m}$ & $1000$\AA  & $100$\AA  \\
\hline
\multicolumn{4}{@{}p{120mm}}{}
\end{tabular*}
\end{table}

The existence of the Abrikosov vortex lattice for conventional
superconductors has been demonstrated experimentally many years
ago through various techniques that have now been applied to study vortex
states in high-T$_c$ superconductors\ \cite{Gammel87,Dolan89}. The lattice
can be imaged in real
space through the Bitter decoration technique in which
fine magnetic dust shows the location of flux lines
as they emerge from the surface of the superconductor. Magnetic neutron
scattering off $\B(\r)$ due to the flux line lattice is another
technique that has been successfully used.  Figure\ \ref{fig3}
clearly shows an image of flux pattern on the surface of $YBCO$
at $T=4.2 K$. A $20$ Gauss field corresponding to $1$ vortex$/\mu
m^2$ was used to obtain this Bitter pattern \ \cite{BolleGammel}.
Figure\ \ref{fig4} shows the reciprocal-space image of the flux line lattice in
$NbSe2$, a strongly type-II layered low-T$_c\approx 7$K superconductor.
The image was obtained via small angle $\approx 1^o$ neutron
scattering with the applied field of $8$ KGauss at $5.2$K.\
\cite{neutronGammel}

\begin{figure}
\caption{Bitter pattern showing
flux lines as they emerge from the surface of $YBCO$, at $T=4.2$K. The
applied magnetic field $H_\perp=20$ Gauss corresponds to $1\mu$m separation
between vortices.  The very straight, somewhat smeared lines cutting across
the lower two corners of the image are twin planes in the YBCO crystal.  These
twins locally orient the Abrikosov flux lattice.  However, since the twins
run at right angles to one another, the two resulting orientations of the
hexagonal
flux lattice are incompatible.  This results in the grain boundary in the
flux lattice which runs from the lower center of the picture toward
the upper left corner.}
\label{fig3}
\end{figure}
\begin{figure}
\caption{Reciprocal space image of the Abrikosov flux lattice, obtained using
neutron
small angle ($\sim 1^o$) scattering from $Nb Se_2$, a strongly type II
low-T$_c$ superconductor ($T_c\approx7$K). The image was
obtained with
$8$ KGauss magnetic field applied along the $c$-axis, at $T=5.2$K.  The six
hexagonally placed small spots are the magnetic Bragg scattering from the flux
lattice.
The large spot at the center is a nonmagnetic background that includes small
angle
scattering from imperfections in the sample.  The scattering at the
higher-order
Bragg spots is too weak to see under these conditions.}
\label{fig4}
\end{figure}

Also, recently, new revolutionary electron holographic techniques have
been used to image the motion of flux lines in thin Pb films. This probe,
which can image flux lines in real time, can be used to study mixed
states also in high-T$_c$ superconductors, and can provide valuable
information about the statics and dynamics of vortex lines\ \cite{Hitachi92}.

Because we are interested in high-T$_c$ superconductors we will
consider strongly type II materials, $\lambda\gg\xi$, which means
that $\A(\r)$ is much ``stiffer'' than $\psi(\r)$, corresponding to
low carrier density or large effective mass. For $H\gg H_{c1}$
the $\B$-field of a vortex overlaps many other vortices,
and one can often make a uniform field approximation,
$\B(\r)=\B=$constant, and concentrate on the dynamics and variations
of $\psi(\r)$ only. This regime can be formally obtained by
taking limits $e\rightarrow 0$ ($\phi_0\rightarrow\infty$) and
$H\rightarrow\infty$ with fixed vortex density
$n=\langle\B\rangle/\phi_0$. This limit corresponds to
uncharged rotating superfluid $^4He$.

\subsection{Mean-field dynamics}

The mean-field dynamics within the infinite penetration length
approximation is assumed to be accurately described by
the time-dependent Ginzburg-Landau theory,
\beqn\label{GLdynamics}
{\pt\psi(\r,t)\o\pt t}={i 2 e
V(\r,t)\o\hbar}\psi(\r,t)-\Gamma{\delta F_{GL}\o\delta\psi^*(\r,t)}\;,
\eeqn
where for simplicity we have left out the Hall effect contribution.
The first term in (\ref{GLdynamics}) is the Josephson term which
leads to ``rotation'' of $\psi(\r,t)$ in the complex plane as the
voltage gradient $\vec {\cal E}(\r,t)=\vec\nabla V(\r,t)$ accelerates the
supercurrent
$\vec j_s(\r,t)=\delta F_{GL}/\delta\A(\r,t)$
\beqna\label{eq:supercurrent}
&\vec j_s&=Re\left[{2 e\hbar \o m}\psi^*(-i\vec\nabla-{2 e\o\hbar
c}\A)\psi\right]\;,\\
&&=\rho_s(\vec\nabla\phi-{2 e\o\hbar c}\A)\;.
\eeqna
We observe that the supercurrent is proportional to the net phase
gradient. In the spirit of the two-fluid model of superfluids the
normal current is simply,
\beqn\label{eq:jNormal}
\vec j_n(\r,t)=\sigma_n{\vec{\cal E}}(\r,t)\;,
\eeqn
where $m$ and $\sigma_n$ must be replaced by their tensor
equivalents for transport in an arbitrary direction,
reflecting the intrinsic material anisotropy\ \cite{Kogan81,commentIsotropic}.

In the Meissner phase there are no vortices present,
and therefore the phase is a true linear superconductor below
critical current density $j_c$. The transport current is however
confined within a penetration length of the surface,
$\vec j_s(\r)\sim e^{-r/\lambda}$, which limits the potential applications.
For $j_s>j_c$ the resulting surface $\B$ field exceeds the
local $H_{c1}$, vortices begin to enter the sample. The ensuing vortex
motion (in response the current-induced forces)
generates dissipation and breaks down superconductivity as we
explain in more detail below for the Abrikosov phase.

To understand transport properties of the superconductor in the
presence of vortices we must first examine the interaction
between the supercurrent and the vortex line. The nature
of this interaction is still a topic of much current
research \ \cite{Thouless93}. It is generally
believed that there are two sources of interaction. One
is the Lorentz interaction of the supercurrent $j_s$ with
the magnetic flux carried by the vortex line, resulting in the force
$\f_L={\phi_0\o c} \vec\tau(z)\times\vec j$, where $\vec\tau(z)$ is a
tangent unit vector to the vortex line. Since we are working in the
large magnetic penetration length regime of nearly uniform $\B$, the
Lorentz interaction should not be very important. A more significant
contribution in this regime is the interaction of the transport
current with the screening supercurrents around the vortex. As is
illustrated in Fig.\ \ref{fig5}, for the counterclockwise supercurrent,
currents add below and subtract above the vortex resulting in higher
and lower kinetic
energy, respectively. The vortex therefore experiences a Magnus
force (analogous to the force that gives an airplane its lift) $\f_M
={\phi_0\o c} \vec\tau(z)\times\j$ which moves it perpendicular to
$\j$, thereby reducing the kinetic energy and the phase difference.

\begin{figure}
\caption{Illustration of the origin of the Magnus force, that the
transport current $\j$ (shown flowing in the $xy$-plane) exerts on the vortex
line emerging from the plane.}
\label{fig5}
\end{figure}

The evolution of the net phase difference between two points
can be obtained from the dynamic Ginzburg-Landau equation
(\ref{GLdynamics}) by assuming that the amplitude of $\psi(\r)$ is
constant,
\beqn\label{eq:PhaseDynamics}
{d\Delta\phi\o d t}= {2 e\o\hbar}\Delta V - 2\pi\times(\mbox{net rate
at which vortices pass between the points})\;.
\eeqn
To maintain steady state therefore requires an applied voltage
difference,
\beqn\label{eq:JosephsonEff}
\Delta V={h\o 2 e}\times(\mbox{net rate at which vortices pass by})\;,
\eeqn
which is an expression of the Josephson effect.

The vortex motion in response to the {\it uniform} transport
current is viscously damped by interaction of normal electrons
in the vortex core with the ions, which in turn results in
dissipation. To maintain d.c. current an electric field ${\cal\E}$ needs
to be applied proportional to the components of the transport
current perpendicular to the vortices,
\beqn\label{eq:BardeenStephen}
{\cal\E}={|\langle\B\rangle|
\o H_{c2}}\rho_n \vec j_\perp\;,
\eeqn
where above proportionality constant is the Bardeen-Stephen flux
flow resistivity\ \cite{BardeenStephen69}. ${|\langle\B\rangle|\o
H_{c2}}$ can be loosely interpreted as the ratio of normal to
superconducting carriers, with only the normal ones contributing the
normal-state resistivity $\rho_n$, in the spirit of the two-fluid
model of superfluids.

Therefore for {\it uniform} current the transport is Ohmic
perpendicular to vortices and is superconducting parallel to
perfectly straight vortices (no thermal
fluctuations)\ \cite{Clem77,Jcomment}.

The situation is quite a bit different for a nonuniform
current\ \cite{HuseWortis93};
a configuration that is more physical since contacts and sample
geometry that produce only uniform current are difficult to achieve.
Consider the linear response of a vortex lattice (an elastic solid in
a viscous medium) to a steady-state d.c. currents, neglecting surface
effects. The steady state motion is determined only by the {\it
total} force and torque. If the current is {\it nonuniform} and if
the currents are fully balanced, the total force and torque vanish.
Then the lattice is simply distorted by the current but no net steady
vortex motion takes place. An example is depicted in Fig.\ \ref{fig6}.

\begin{figure}
\caption{Illustration of vortex lattice distortion due to the presence of
non-uniform currents. As described in the text, since the total force and
torque vanish, the lattice distorts but does not move as a whole. This
therefore illustrates the vanishing of linear resistivity at finite
wavevectors.}
\label{fig6}
\end{figure}

We can resolve $\vec j(\r)$ into Fourier modes,
\beqn\label{eq:Fourier}
\vec j(\r)=\int d^3 r e^{i\k\cdot\r}\; {\vec{\hat j}}(\k)\;,
\eeqn
and ask about the electric field response due to such a plane wave of
current,
\beqn\label{eq:Ek}
{\hat{\cal E}}_j(\k)=\hat{\rho}_{ij}(\k){\hat j}_i(\k)\;,
\eeqn
where the summation over spatial indices is implied.

As we already mentioned for the current parallel to the vortices,
\beqn\label{eq:Rholong}
\hat{\rho}^{||}_{ij}(\k)=0\;.
\eeqn
The current perpendicular to the vortices couples to ``phonon''
modes of the vortex lattice, imposing {\it static} elastic distortions
but no steady motion for $\k\neq\vec 0$. Therefore we obtain a
highly {\it nonlocal} resistivity with ${\cal{\E}}(\r)$ determined by
$\vec j(\r')$ over the entire sample.
\beqn\label{eq:Rhoperp}
\hat{\rho}^{\perp}_{ij}(\k)\sim\delta(\k)\;.
\eeqn
We therefore conclude that in the absence of fluctuations, the vortex
lattice superconducts {\it except} for $\k=0$ and
currents perpendicular to the vortices. Since in the normal state the
resistivity is fully local, dissipating at all wavevectors $\k$,
$\hat{\rho}_{ij}(\k)=\rho_n$, in mean-field theory $\rho_{ij}$ is
discontinuous at the NS transition (at $H_{c2}(T)$).

\section{THERMAL FLUCTUATIONS}

We now examine how the mean-field results discussed in the previous
sections are modified by thermal fluctuations. In low temperature
superconductors these fluctuations are not very important except
very close to phase boundaries and in thin films or wires. The small size
of the critical region in bulk low-T$_c$ materials is associated with the fact
that the thermal length
\beqn\label{LambdaT}
\Lambda_T={\phi_0^2\o 16\pi^2\T}\approx {2\times 10^8 {\rm\AA K}\o T}\;,
\eeqn
is much larger than any other length in the problem except very
close to $T_c$ where other lengths diverge.

For the high-T$_c$ superconductors simple
considerations also allow us to construct the ``Ginzburg criterion'',
which determines when the effects of thermal fluctuations become
important. This occurs when the ordering condensation
energy in a coherence volume is equal to the thermal disordering
energy, $\alpha^2\xi^2\xi_z/2\beta\approx\T$.  Stated equivalently
in terms of material parameters,
\beqn\label{Ginzburg}
\lambda\approx{\gamma\phi_0^2\o 48\pi^3\T\kappa}\Lambda_T=
{\gamma\Lambda_T\o 4\pi\kappa}\;.
\eeqn
We therefore observe that for the highly anisotropic
high-T$_c$ superconductors the
effective thermal length is reduced by more than $4$ orders of
magnitude by the combination of high T$_c$, large anisotropy
$\gamma^{-1}\approx 10^2$ and large $\kappa\approx 10^2$. This
therefore greatly expands the critical region in which mean-field
theory no longer applies and fluctuation-corrected theory must be used.

We first consider fluctuations in the ideal case of disorder-free
superconductors and then go on to study the effects of disorder at
finite temperatures.

\subsection{Disorder-free samples}

The finite temperature effects can be incorporated into the
statics by summing over all the configurations of the system
weighted by Boltzmann weight $e^{-F_{GL}/\T}$, instead of just
simply minimizing the Ginzburg-Landau free energy $F_{GL}[\psi,\A]$,
Eq.\ref{eq:FreeEnergy}. The effects of thermal fluctuations can
equivalently be incorporated in the dynamic Ginzburg-Landau equation
(\ref{GLdynamics}) by the addition of white noise $\zeta(\r,t)$
of intensity proportional to $\T$, as set by the
fluctuation-dissipation theorem.

Near the phase boundaries the fluctuations are most effective
in disrupting the ordered phases, because there the stiffness of the
order parameter vanishes. For example, for the Meissner phase
$\rho_s^{MF}(T)\sim(T_c^{MF}-T)$. Similarly, for the vortex lattice
phase the shear modulus plays the role of the stiffness, which
vanishes near the $H_{c2}^{MF}$ as
\beqn\label{Shear}
\mu(H)\sim {(H_{c2}^{MF}-H)^2\o\lambda^2}\;.
\eeqn

The idea that the flux line lattice in high-T$_c$ materials
might melt into
a liquid ($\mu=0$) of freely moving vortex lines was first
suggested by Nelson\ \cite{NelsonSeung89}. Recently there has been much
experimental evidence that in fact a large portion of the
Abrikosov lattice is replace by the vortex liquid, all the way up to the
mean-field $H_{c2}^{MF}$ boundary where the superconductor
crosses over to normal behavior\ \cite{Gammel88} (see figures\ \ref{fig1}(b),
\ref{fig7}). Although a full theory of the
melting transition does not exist, an estimate of the location of the
melting boundary can be made using the Lindemann criterion. Much work
has been directed to accurately describe the melting boundary,
carefully taking into account the nonlocality of the elastic moduli
of the vortex lattice\ \cite{Houghton89,Brandt89}. However to avoid drowning
in technical details we illustrate the idea ignoring the wavevector
dependence of the elastic moduli.

As is true for real ionic lattices, the phonon modes of the
vortex lattice are thermally excited with mean elastic energy
$\T/2$. The mean square displacement of a vortex is a rough
measure of the disruption of crystalline order by thermal
fluctuations,
\beqna\label{eq:Displacement}
&\langle|\vec u(\r,t)|^2\rangle&=
\int d^3 k\langle|{\vec{\hat u}}(\k,t)|^2\rangle\;,\\
&&\sim\int_0^{k_\perp\sim 1/a, k_z\sim 1/\xi_z}
d^3 k{\T\o\mu k^2}\;,\\
&&\sim{\lambda^2\T (H_{c2}^{MF})^2\o\xi_z (H_{c2}^{MF}-H)^2}\;.
\eeqna
As expected the displacement increases with temperature and as
$H_{c2}^{MF}$ is approached.  Melting occurs when
the displacement becomes some significant fraction of the lattice
spacing $a$,
\beqn\label{Ratio}
\langle|\vec u(\r,t)|^2\rangle^{1/2}\approx c_L a\;,
\eeqn
where $c_L\approx 0.1$ is the empirically determined Lindemann ratio.

The resulting vortex liquid has local short-scale superconductivity,
but the highly mobile disordered vortices do not allow global phase
coherence to set in.  In this sense the vortex
liquid is the same disordered thermodynamic phase as the normal state,
exhibiting only the quantitative difference of having much larger
fluctuations of the superconducting order parameter.
In both of these normal-state regimes $\langle\psi\rangle=0$, unlike in the
truly superconducting Abrikosov lattice and Meissner phases.
At $H_{c2}^{MF}$ a gradual crossover occurs as well-
defined vortices form, and conductivity and diamagnetism increase
due to the local superconductivity.  However, no thermodynamic singularity
occurs since no symmetry is broken as $H^{MF}_{c2}$ is crossed.

The interaction between flux lines becomes very weak and falls off
exponentially for separations $r>\lambda$. Recently it has
been argued that because of the weak interactions, at weak
fields a sliver of vortex liquid will also intrude right above
$H_{c1}$\ \cite{NelsonSeung89,FFH89}, as is illustrated in Fig.\ \ref{fig1}(b).
However the verification of the existence of the flux-line liquid in
this portion of phase diagram has up to now alluded experimentalists.

A variety of liquid phases have been suggested such as the
entangled and disentangled vortex liquid\ \cite{NelsonSeung89}
as well as the intermediate hexatic phase\ \cite{Marchetti90,Worthington90}
characterized by the quasi-long-range orientational order and
short-range translational order in analogy with the $2d$-melting
theory\ \cite{NelsonHalperin79}.
Recently, there have also been suggestions that in layered
superconductors there is a possibility of two thermodynamically
distict vortex lattice phases. One is the conventional $3d$
Abrikosov lattice with vanishing linear resistivity. The second
is a quasi-$2d$ vortex lattice phase, with finite linear resistivity,
also known as the supersolid phase\ \cite{GlazmannKoshelew90,Frey93}.
In Fig.\ \ref{fig7} we have presented a schematic phase diagram showing the
expected location of these novel vortex phases.  In
the $\langle B(\r)\rangle$ versus $T$
phase diagram of Fig.\ \ref{fig7}, the Meissner phase is collapsed
down to the $T$-axis.
The nature of the novel hexatic and entangled liquid phases has been
extensively discussed\ \cite{NelsonSeung89,Marchetti90,Worthington90}.
We now briefly describe the less known putative novel quasi-$2d$ lattice
phase.

\begin{figure}
\caption{Some of the various newly proposed vortex phases are illustrated. The
quasi-2d-supersolid vortex lattice phase is expected to be a distinct phase
from the conventional Abrikosov 3d-lattice, and will exist at high
$B$-fields, where the inter-planar coupling is weak.}
\label{fig7}
\end{figure}

In a usual molecular crystal an interstitial is a point defect that
costs a finite amount of energy $E_i$ and in equilibrium exists at
densities $\sim e^{-E_i/\T}$. On the other hand in a three-dimensional vortex
lattice a vortex cannot end inside a superconductor (except on a
magnetic monopole, which is highly unlikely) so an interstitial
is a {\it line} defect with a constant free energy per unit of length. Its
free energy is therefore proportional to the thickness of the
sample. Since in the thermodynamic limit this defect energy
diverges, at equilibrium we expect no interstitials in the
vortex lattice and magnetic flux per unit cell is exactly
$\phi_0$. This situation is to be contrasted with thin film
superconductors where the interstitials and vacancies are
{\it point} defects, of finite energy and therefore are present in
finite density at finite temperature. Concomitantly the flux per
unit cell does {\it not} equal to $\phi_0$, and hence the number of
flux points per unit cell fluctuates as the interstitial and
vacancies move around.

The crossover between the $2d$ and $3d$ behavior in layered $CuO_2$
superconductors is described by the Lawrence-Doniach
model\ \cite{LawrenceDoniach71} with free energy,
\beqn\label{LDmodel}
F=\sum_i\left\{F_{2d}\left[\{\psi_i(\r)\}\right]-\int d^2 r {\rm
Re}\left[J\psi^*_i(\r)\psi_{i+1}(\r)\right]\right\}\;,
\eeqn
where we have taken $\B$ to be perpendicular to layers. The first
term is the free energy of independent $2d$ layers labeled by index
$i$, and the second term gives local coupling between adjacent
layers.

At $J=0$, the $2d$ uncoupled layers allow for vacancies and the
interstitials with entropy dominating over energy. As the coupling
is increased at some critical value $J_c=J(T_c)$ the interstitial free
energy per unit of length (tension) becomes positive, the system
makes a transition to a $3d$-dominated behavior and the net
interstitial density vanishes. Since the effective interlayer
coupling is actually $\hat{J}=J a^2\sim J/\langle B\rangle$ the system
becomes more $2d$-like at higher fields. At large fields (small $\hat{J}$)
the interstitials form a highly flexible and mobile line-liquid
existing inside a $3d$ vortex lattice with a true crystalline order. Since
the interstitials are free to move, this supersolid is not
superconducting for the reasons explained in previous sections.
The resulting phase diagram is illustrated in Fig.\ \ref{fig7} where
we have plotted $\langle B\rangle$ versus $T$.

\subsection{Nonlocal resistivity at finite temperature}

As we already discussed in Sec.2.2, the transverse flux-flow
resistivity of a vortex lattice is highly nonlocal,
\beqn\label{Nonlocal}
{\cal\E}(\r)=\int d^3 r'\rho(\r-\r')\vec j(\r')\;,
\eeqn
and in fact in the absence of plastic flow (possible in very
clean samples)
we expect $\rho^{ab}_{lattice}(\r)\sim 1/{\rm volume}$. This form is
to be contrasted with the completely local resistivity in the normal
state, $\rho_{normal}\sim\delta^{(3)}(\r)$.  In the presence of
thermal fluctuations the flux-line lattice melts over a large
portion of the HT phase diagram. The resulting vortex-line liquid
consists of flexible, highly mobile flux lines that can easily
respond to the transport current and therefore generate finite flux-flow
resistivity. Unlike the vortex lattice, in the vortex/flux-line liquid phase
there is no perfect correlation in the motion between different, separated
vortex lines. As the lines traverse the sample along the
applied magnetic field (say in the $z$-direction) in a distance $z$
they will wander transversely a distance $(z\T/\tilde{\eps})^{1/2}$,
where the line
tension $\tilde{\eps}=\phi_0 H_{c1}/4\pi$.  Even in thin samples
a typical line will deviate away from its average direction
much further than the average interline separation set by
the applied magnetic field (see Table \ref{table1}).  Therefore, the flux
lines experience many collision with their neighbors entangling with
many distant ones, much like directed polymers in a solution.
Although no rigorous calculation of the flux-line crossing barrier
energy is available, simple estimates give $E_x\approx
50\T$\ \cite{NelsonSeung89}, far away
from $H_{c2}$, where it is expected to vanish as
$\sim\ln[H^{MF}_{c2}/H]/\ln[H^{MF}_{c2}/H_{c1}]$\ \cite{Obukhov90}. We
expect that these frequent flux-line interactions and
entanglements will lead to a large transverse viscosity of the
vortex-line liquid\ \cite{Radzihovsky93} which replaces the perfect
correlations controlled by the finite shear modulus in the Abrikosov
lattice. Besides these transverse viscous correlations, the
connectivity of the vortex lines leads to an even stronger
nonlocality in the viscosity of the line-liquid along the direction of
the applied field. Since the
viscosity is a measure of the response of velocity to an
applied force, it is directly related to the flux-flow resistivity.
The flux-flow resistivity of vortex liquid should therefore exhibit
a nonlocal behavior in-between that of a completely local one of the normal
state and a completely nonlocal one of the vortex lattice. As the
superconductor is cooled
into the vortex liquid, $\rho(\r)$ becomes more and more extended,
especially parallel to vortices.

The nonlocality of the resistivity in the liquid phase
has been recently studied experimentally\
\cite{SafarNonlocalJ92,Busch92,SafarNonlocalJ93} and
theoretically\ \cite{HuseMajumdar93}
in the geometry depicted in Fig.\ \ref{fig8}. The transport current
at the top surface of the superconductor exerts a transverse
force on top ends of vortex lines.  Since the vortex lines are well
connected and
rarely break, the current at the top of the sample pulls on the
whole line. This results in the vortex motion and therefore electric field
far away from the region where the current actually flows (top).
Since the voltmeter
measures the average rate at which vortex lines pass between the two
contacts, excluding the possibility of line breaking we have
$V_{top}=V_{bottom}$, with the vortex velocity proportional to the
total current integrated from the top to the bottom of the sample.
This situation is to be
contrasted with the normal state local resistivity where the
$V_{top}>V_{bottom}$ because current density is higher at the top.
In experiments of Safar, et al.\ \cite{SafarNonlocalJ93} on $35\mu m$
thick $Y_1Ba_2Cu_3O_7$ crystal, in a regime where vortices have a good
integrity over a distances larger than the sample thickness, the effect
of nonlocal resistivity has been observed.

\begin{figure}
\caption{Schematic of the experimental setup to measure nonlocal
resistivity of the vortex liquid phase. In contrast to the normal state,
transport in the vortex liquid will generate $V_{top}=V_{bottom}$.}
\label{fig8}
\end{figure}

Supposing that the Fourier transform of the conductivity is an
analytical function of $\k=(\k_\perp,k_z)$ for small $k$,\
\cite{AnalytComment},
nonlocal resistivity of the
vortex liquid can be phenomenologically described
as \ \cite{Marchetti91,Radzihovsky93}
\beqn\label{PhenomRho}
\rho(\k)\sim{1\o \gamma+\eta_{ab}|\k_\perp|^2 + \eta_z k_z^2}\;,
\eeqn
where $\gamma$ is the local friction (related to the local conductivity of
Bardeen and Stephen), and $\eta_{ab}$, $\eta_z$ are the
vortex liquid viscosities leading to nonlocality in the flux-flow
resistivity. This model can be used to solve for current and voltage
patterns in a realistic experimental geometry by solving $4$th order
partial differential equations.

\subsection{Static disorder}

In real high-temperature superconductors disorder plays an essential role.
As we have already briefly described, it modifies the HT phase diagram, but
more importantly, the disorder tends to pin vortices in place, drastically
reducing the flux-flow resistivity, as we now describe in more detail.

The most important contribution of static disorder leads to a
spatially varying transition temperature, $T_c(\r)$.
The disorder effects can therefore be incorporated into
the Ginzburg-Landau description by allowing the $\alpha$
parameter to be a function of position,
\beqn\label{randomAlpha}
\alpha(\r,T)\approx a(\r)\left(T-T^{MF}_c(\r)\right)\;,
\eeqn
where the properties of the function depend on the nature of
disorder. Chemical and structural imperfections in the material
naturally lead to random $\alpha(\r)$ with only short-range
correlations. Correlated random static disorder is also possible and
might arise from grain or twin boundaries, screw dislocations,
artificially created ion columnar tracks\ \cite{Civale91,Budhani92}, or
from an epitaxial multilayers and e-beam written
patterns.  Finally the crystal structure itself, such as for
example the $Cu O_2$ layers of the high-T$_c$
superconductors, provides a natural modulation of $T_c(\r)$
describable by a periodic $\alpha(\r)$.

Since the core of a vortex line is in a normal state, the energy is
minimized when the vortex is located in the region of weakest
superconductivity, where $\alpha(\r)$ is greatest and minimum amount of
condensation energy is lost. This leads to pinning of a vortex line by normal
impurities, dislocations, grain/twin boundaries and to the
intrinsic pinning in between the $Cu O_2$ planes
(where the superconductivity is weakest).

The phase of the wavefunction cannot be measured and therefore the
disorder cannot directly couple to the phase $\phi(\r)$. We
therefore conclude that the Meissner phase is stable to disorder,
except near its phase boundaries.

Both the strong and weak pinning limits have been previously studied.
In the strong pinning limit the pinning energy  exceeds the
vortex-vortex interaction energy and leads to a fully disrupted
vortex lattice. The Ginzburg-Landau free energy is minimized by a
random complex function $\psi(\r)=\psi_0(\r)e^{i\phi(\r)}$.

It has been known for some time that the Abrikosov phase is unstable
to introduction of disorder\ \cite{Larkin79} and the long-range
translational order is destroyed beyond a disorder-determined
Larkin-Ovchinnikov length $L_{LO}$.  It has been recently proposed that the
resulting new low temperatures and fields phase possesses
long-range order with a sample and disorder-specific random vortex pattern.
For the case of point (uncorrelated) disorder the equilibrium phase is the
vortex glass with the name chosen by the analogy with spin glass, where
the ground state order parameter is random but frozen with long-time
correlations\ \cite{Shih84,FFH89}. For correlated disorder such as
columnar pins, an anisotropic version of the vortex glass has been proposed
and dubbed Bose glass\ \cite{NelsonVinokur92} by the isomorphism
of the resulting theory with bosons on random
substrate\ \cite{FWGF89}. Here we will confine our review to the
isotropic vortex glass, with the analysis easily extendible to
Bose glass.

The stability of the vortex glass phase to thermal fluctuations
depends on the nature of the low-lying states. By analogy with the
scaling theory of spin glasses\ \cite{FisherHuse88}, the excitations
are ``bubbles'' inside a ground state and correspond to the displacement
of one or more vortices a distance $L$ in a region of
linear size $L$. The average energy of the low-lying excitation is
expected to scale with a power of $L$
\beqn\label{Eexcite}
E_{e}\sim\Upsilon L^\theta\;.
\eeqn
Since for a circular loop, $\theta_l=1$, characteristic of excitations in
the Meissner phase, we expect that $\theta\leq 1$.

The stability of the vortex glass phase clearly depends on the sign
of the $\theta$ exponent. If $\theta<0$ then the large-scale
excitations cost very little energy and will be present at $T>0$ at
equilibrium. This turns out to be the case in two dimensions
($d=2$). The correlation length then scales as a power law with
temperature
\beqn\label{VGcorrLength}
\xi_{VG}\sim T^{-1/|\theta|}\;,
\eeqn
with the long-range order only present strictly at $T=0$.

In three dimensions no satisfactory analytical analysis is
available to date, primarily because of the difficulty associated
with treating the dislocations (although some arguments have
proposed that the dislocations might be irrelevant). Numerical
studies of highly simplified models treat the 3d vortex glass phase
in the limit $\lambda\rightarrow\infty$ and find
$0\leq\theta\leq 0.3$\ \cite{Reger91,Cieplak,Gingras92}. These works point
to the stability of the vortex glass phase in $d=3$, for $T>0$.

The stability of the vortex glass phase in three dimensions to
thermal fluctuations in the presence of screening ($\lambda$
finite) is still an open question. Since the interactions
crucial for the formation of the vortex glass phase become
exponentially weak for $r>\lambda$,
it appears that screening tends to weaken the
stability of the vortex glass.  Experimentally this regime is only
probed when $\xi_{VG}>\lambda$ which corresponds to resistivities
well below those studied in experiments to date.

In the case of weak disorder a topological glass phase might
result, with properties intermediate between that of the vortex
lattice and the vortex glass. As we have seen vortex glass
is a phase highly correlated in time, but spatially it looks like a frozen
vortex liquid with many dislocations present. In the vortex
glass the {\it strong} disorder destroys the
topological order and dislocations proliferate. On the other hand,
in the absence of disorder,
the vortex lattice has both positional and orientational
long-range order and is therefore dislocation-free at large scales.
It is possible that in the presence of {\it weak} disorder an
intermediate topological glass forms in which the topological order
of the vortex lattice survives.

The resulting topological glass phase is then describable as a
dislocation-free selastic medium (lattice) in a random potential. The
displacement of a vortex from a position $\r=(\x,z)$ due to disorder
is described by $\u(\r)\perp{\hat z}$, with vortices parallel to
$\hat z$. The Hamiltonian in $d$ dimensions is then given by
\beqna\label{HamiltonianGlass}
H&=&\int d^d r\; {\vec\nabla}\u(\r)\cdot\stackrel{\leftrightarrow}{\bf E}
\cdot{\vec\nabla}\u(\r)\nonumber\\
&&+\sum_i\int dz\; V\left(\{\x_i+\u(\x_i,z)\},z\right)\;,
\eeqna
where $\x_i$ is the $d-1$ dimensional position of the $i$-th line in
the space transverse to $\hat z$ and the sum is over discrete vortex
lines. The first term, above, is the elastic
energy, quadratic in the displacements $\u(\r)$,
described in terms of the elasticity
tensor $\stackrel{\leftrightarrow}{\bf E}$\ \ \cite{Houghton89,Brandt89}.
The second term is the contribution due to the pinning disorder,
described by a random potential $V(\r)$ with spatial correlations
that depend on the type of disorder.

The ground state without dislocations has been studied analytically
and is characterized by displacement correlations that
are logarithmic in position
\beqn\label{uCorrelations}
{\overline{\left[\u(\r_1)-\u(\r_2)\right]^2}}\sim\ln |\r_1-\r_2|\;,
\eeqn
for large $|\r_1-\r_2|$\ \cite{Villain,GiamarchiLeDoussal93}.
Weak disorder destroys long-range {\it positional} order replacing
the Bragg peaks by power-law singularities characteristic of the
quasi-long-range order. Because of the weakness of the logarithmic growth
of displacements, the dislocation density vanishes and the topological order
remains, consistent with the assumption of the model.
Full long-range {\it orientational} order, however,
still survives in this randomly pinned vortex lattice.

For the topological glass to exist as a distinct phase from the
strongly disordered vortex glass it must be stable to dislocations.
To establish this stability we look at the
low-lying excited states that disrupt the order of the topological
glass. An appropriate low-energy excitation is a patch of size $L$
moved over by one lattice spacing $a$ with respect to the rest
of the vortex lattice. The elastic nonlinear distortion is
$\u_o(\r)$, outside the patch and $\u_i(\r)$ inside the patch, with the
displacement jump $|\u_i(\r_{wall})-\u_o(\r_{wall}+a\hat n)|=a$,
concentrated on the wall bounding the patch (see Fig.\ \ref{fig9}).
The elastic energy is
unchanged in the interior and exterior of the patch.  The average
total energy is increased due to the elastic strain by $a$ and
for optimized wall position scales as
\beqn\label{Eexcite2}
{\overline E}_{e}(L)\sim\Upsilon L^\theta{_w}\;.
\eeqn
Since the elastic energy is concentrated on the wall, the average
total energy can at most scale as the area of the wall, which puts
an upper bound on $\theta_w$,
\beqn\label{AreaBound}
\theta_w\leq 2\;,
\eeqn
where the optimization of the wall position can lower $\theta_w$ below
$2$.

\begin{figure}
\caption{Illustration of a domain wall of nonlinear distortion
resulting from shifting a patch of topological vortex glass by a lattice
vector $a$. The single  valued vortex displacement $\u_i(\r)$ inside the
wall differ from the distortions outside the patch, $\u_o(\r)$ by $a$. The
displacement field jump of $a$ is concentrated on the domain wall.}
\label{fig9}
\end{figure}

Although the total average energy of the wall is positive,
large sections of the wall can have a negative energy that scales
in the same way as the average energy in Eq.\ref{Eexcite2}.
A section of the wall with negative energy can be created by an
addition of a dislocation line, as illustrated in Fig.\ \ref{fig10}.
Upon encircling the dislocation
line $\u_i-\u_0$ must change by a Burger's vector $\bf b$,
a change that is concentrated on the domain wall. Therefore the
energy of adding a dislocation equals to the local core energy plus
the wall energy. On scale $L$ the system might prefer to create
a dislocation line with the negative energy domain wall compensating
the core energy cost
\beqn\label{Edislocation}
E_{dislocation}\sim L - \Upsilon L^\theta{_w}\;.
\eeqn
\begin{figure}
\caption{An addition of a dislocation line together with a
negative energy domain wall might lower the energy of
the topological glass, depending on the value of the $\theta_w$ exponent.
This would result in distablization of the topological glass toward a
fully disorder vortex glass.}
\label{fig10}
\end{figure}

Hence we expect that the system can lower its energy by adding
dislocations on large scales if $\theta_w>1$ and the
topological glass phase will always be unstable to the vortex glass without
topological order. On the other hand if $0<\theta_w<1$ the topological
glass phase should be stable against dislocations for weak disorder.
To date the precise range into which $\theta_w$ falls, and therefore the
stability of the topologically ordered, dislocation-free
vortex glass phase is an open question.

\section{NONLINEAR RESISTIVITY IN SUPERCONDUCTING PHASES}

Vanishing linear resistivity is probably the most natural and
practical distinction between a superconducting and normal phase.
As we have already seen, the vortex liquid has nonlocal but
nonvanishing linear resistivity.  On the other hand a vortex lattice
can be easily pinned with few strong inhomogeneities or
even naturally pinned by the boundaries. In this pinned
state we have seen that the vortex lattice has a true
superconducting response to a {\it uniform} current.

Conventional theory of transport in disordered vortex states, based on
the ideas of Anderson and Kim\ \cite{AndersonKim64}, describes the
resistivity in terms of thermally activated motion of independent
vortex bundles over the finite energy barriers $V_p$, introduced by the
disorder\ \cite{Kes89,Brandt91}. The size of the bundles is assumed to be
finite, on the order of Larkin-Ovchinnikov length $L_{LO}$, beyond which the
lattice order is destroyed by disorder\ \cite{Larkin79}.
The transport current introduces an average tilt to the random landscape
potential and therefore an overall thermal drift of vortex bundles in the
direction of the tilt. The vortex drift destroys the phase coherence of
superconducting wavefunction, resulting in finite vortex-flow
{\it linear} resistivity of Arrhenius form, $\rho_{l}\sim\exp(-V_p/\T)$.
Therefore this conventional theory maintains that at finite temperatures,
the disordered low temperature vortex phase is not really superconducting,
so is qualitatively identical to the non-superconducting vortex liquid regime.

The difference between the conventional picture and the
transport in the vortex glass phase comes from the strong
inter-vortex interactions not included in the Anderson-Kim picture.
Upon cooling a sharp thermodynamic transition takes place from a
non-superconducting vortex liquid to a distict
superconducting vortex glass phase\ \cite{Koch89}. The vortex motion
in the vortex
glass phase is still described by the thermally activated motion, except
that the strong correlations lead to bundle and therefore barrier sizes that
diverge as $j\rightarrow 0$. We now examine the vortex glass phase along with
Meissner phase, and show that these phases possess only nonlinear resistivity
and are therefore true linear superconductors.

In the Meissner phase it is the
nucleation and growth of vortex loops, and in the vortex glasses
phases it is the collective motion of already present vortices in
response to a transport current that are the excitations that lead to energy
dissipation. Since the motion of a bundle of vortices can be
described by the creation and growth of vortex loops, in both phases
the low-lying excitations are vortex loops superimposed onto the
ground state background of each phase. We can use the scaling theory
of these low-lying excitations to analyze the resistivity both in
the flux-free Meissner phase and the frozen vortex glass phase.

For the well understood Meissner phase we know the nature of the
excited states (vortex loops) and therefore can predict the exponents in
addition to the scaling form of resistivity. Unfortunately, for the vortex
glass neither the ground state nor the low-lying excited states are well
understood. This lack of understanding however, can be neatly packaged into
a set of exponents that enter the scaling theory of the nonlinear resistivity.
Some attempts have been made to apply a collective pinning theory to
calculate these exponents. However, these theories ignore the existence of
dislocations and therefore breakdown of long-range translational order on
length scales $L > L_{LO}$\ \cite{FGLV89}. It is likely that dislocations
will have an affect on the value of these exponents.

As was described in the previous section, the energy of low-lying
excitations scales as $\Upsilon L^\theta$, with $\theta$ being
specific to the type and shape of the excitations and the phase-background
in which they are created. For example for the isotropic
Meissner phase the lowest energy excitations are circular loops
characterized by superfluid density stiffness $\Upsilon\sim\rho_s$
and $\theta=1$. The transport current-generated Magnus force
couples to the area $A$ projected onto the plane perpendicular to
the current
\beqn\label{Area}
A(L)\sim A_0 L^\tau\;.
\eeqn
The exponent $\tau$ has a lower bound of $2$, saturated by
a circular loop in the Meissner phase with $A_0=1$ and $\tau=2$. The energy
supplied by the Magnus force is
\beqna\label{EnergyMagnus}
&E_{c}(L)&\sim f_{M}A(L)\;,\\
&&\sim j L^\tau\;,
\eeqna
where $j$ is the transport current density. Balancing the elastic
energy of the excitation with the current energy we obtain the
average size $L_j$ of the lowest energy excitation
\beqn\label{ExciteL}
L_j\sim j^{-1/(\tau-\theta)}\;,
\eeqn
that are produced when $\tau>\theta$.

In order to reach these low-lying states the system
might have to overcome barriers by moving through less favorable
regions. In general the size of the energy barriers will scale as
\beqn\label{Ebarrier}
E_{b}=\Delta L^\psi\;,
\eeqn
where for loops in the Meissner phase $\psi=\theta=1$. In general,
however, the barriers are usually larger than the final energy
state, and therefore $\theta$ is a lower bound for $\psi$, i.e.
$\psi\geq\theta$. These two energy scales are schematically depicted
in Fig.\ref{fig11}.

\begin{figure}
\caption{Local random potential due to impurity disorder is
characterized by the typical size of its barriers, scaling as
$L^\psi$, and by the energy difference between ground state and next
low-lying state, scaling as $L^\theta$.}
\label{fig11}
\end{figure}

The rate of energy dissipation via the production of the low-lying
excitations is governed by the Arrhenius law $\sim exp{\left[-\Delta
L_j^\psi/\T\right]}$. The resulting nonlinear resistivity due to the
vortex motion over the barriers is
\beqn\label{muExp}
{{\cal E}\o j}=\rho_{nl}(j)\sim\exp\left(-c\over\T j^\mu\right)\;,
\eeqn
where $\mu=\psi/(\tau-\theta)$. The excitations with
the smallest $\mu$ dominate with the upper bound
again set by the vortex loops as in the Meissner phase
where $\mu_{Meissner}=1$. We immediately observe that for
$\tau>\theta$ the barriers diverge as $j\rightarrow 0$ and
the linear dc resistivity vanishes
\beqn\label{LinearR}
\left.{d {\cal{E}}\o d j}\right|_{j\rightarrow 0}\equiv\rho_{l}=0\;.
\eeqn

Another consequence of Eq.\ref{muExp} is a slow decay of
non-equilibrium screening supercurrents that can be detected by measuring
the relaxation of the associated magnetization $M(t)$,
\beqn\label{decayM}
{d M(t)\o d t}\propto {d j(t)\o d t} \propto {\cal E}(j)\;,
\eeqn
where ${\cal E}(j)$ is the field needed to prevent $j(t)$ from decaying.
Combining
(\ref{decayM}) with (\ref{muExp}) leads to a slow logarithmic decay at
long times,
\beqn\label{decayJ}
j(t)\approx j_o \left[\ln(t/t_0)\right]^{-1/\mu}\;,
\eeqn
with $t_0$ as the microscopic time of order $10^{-9} - 10^{-12}$ seconds.

Recent transport experiments on $Y Ba_2 Cu_3 O_7$ thick films
(in 3d regime) find nonlinear resistivity behavior
described above with $\mu\approx0.2-0.3$ which sets in when the temperature
is lowered past a well defined transition temperature $T_{vg}$ (see next
section)\ \cite{Koch89,Gammel91}. Later much more sensitive magnetization decay
experiments on $YBCO$ films have found a scaling behavior in agreement with
Eq.\ref{decayJ}, with $\mu\approx 1/3$\ \cite{Sandvold92,Dekker92}. These
experiments therefore provide confirmation of the vortex glass picture.

\section{PHASE TRANSITIONS}

Having discussed the nature of the phases that occur as a result
of thermal fluctuations and disorder we now turn our discussion to
the phase transitions between these new phases.

\subsection{First-order vortex lattice melting transition in pure crystals}
In the absence of disorder the low temperature and low field
phase is the vortex lattice, which melts into the vortex liquid
upon increasing field and/or temperature. This transition has been
extensively studied experimentally, numerically and analytically.
Transport experiments on clean $Y_1 Ba_2 Cu_3 O_7$ crystals show
an abrupt (with milliKelvin resolution), hysteretic resistance drop
to zero at the melting transition\ \cite{SafarMelt92}. This experimental work
provides solid evidence for the vortex lattice melting transition being
first-order.  The melting transition
has {\it not} yet been detected thermodynamically (i.e. by
measuring latent heat or magnetization jumps) due to smallness of
the vortex latent heat, estimated to be $\leq\T_M$ per vortex per $Cu O$ layer.
The vortices are very dilute and the specific heat due to electrons and
phonons of the underlying solid are quite large at $T_M\approx 70 {\rm K}$
in comparison and dominate over the vortex lines' contribution.
Also since the high-T$_c$ materials are highly
anisotropic and therefore are in the quasi-2d regime, the
first-order transition is expected to be weak.

Evidence from the simulations of
London theory (amplitude of $\psi(\r)$ is a fixed constant) on a lattice
also clearly points toward the first-order
transition\ \cite{Hetzel92}.
Further evidence is provided by the analytical work of Brezin,
{\it et al.}\ \cite{Brezin85}. There an $\epsilon=6-d$ expansion is
carried out which shows that the mean-field theory
second-order transition at $H^{MF}_{c2}$ is destabilized by fluctuations.
The resulting runaway of renormalization-group flow is taken as an
indication of fluctuation driven first-order
transition\ \cite{FluctComment}.
Finally based on general symmetry grounds Landau theory for melting
has a term cubic in the order parameter and therefore leads to
first-order melting transition in agreement with usual melting of 3d
crystals\ \cite{seeAppendix}.

\subsection{Stability of first-order transition to disorder}

We have previously described the strong influence disorder has on
vortex phases, for example converting
the vortex lattice to a vortex glass.  We now examine the stability
of the first-order melting transition to weak disorder using
Imry-Wortis general scaling arguments\ \cite{ImryWortis79}.

Near the melting transition, disorder's local and random preference
for the solid (vortex glass) or vortex liquid phase can be described
via Landau theory with a random local variations in $T_M$
(see Eq.\ref{randomAlpha}). An increase in free energy due to a
creation of an island of one phase inside the other will have a
positive interface contribution and a favorable negative bulk
contribution due to the disorder
\beqn\label{Island}
\delta F\sim \sigma L^{d-1} - \Delta L^{d/2}\;,
\eeqn
where $\sigma$ is the tension of the $d-1$-dimensional interface and
$\Delta$ is the disorder strength.  The negative disorder
contribution is a sum of random energies (with zero mean) and
therefore by central limit theorem scales as a square root of number
of impurities in the region of size $L$.  For $d>2$ and weak
disorder (small $\Delta$), the free energy of the island is positive and their
density will be strongly suppressed.
Therefore, there will be a stable two-phase coexistence with the phase
transition remaining first-order. On the other hand for $d\leq2$ the
disorder contribution will dominate the interfacial energy and
system will create interpenetrating islands of one phase inside the
other. These arguments therefore suggest that the melting transition
of the vortex glass into a vortex liquid will become a continuous
transition as the effective dimensionality decreases and the strength of
disorder increases.

Recent experiments on untwinned, single crystal $Y Ba_2 Cu_3 O_7$
by Safar and coworkers\ \cite{SafarMelt93} find a behavior that can be
reconciled with the above theoretical picture. They find a
hysteretic, abrupt, first-order transition at weak applied fields
$H$ which becomes a continuous transition at higher $H$. It is also
observed that the first-order regime increases as pinning decreases
with the tricritical point moving in a range $5-15$ Tesla. Because
increasing $H$ is equivalent to lowering $T_m$ and making the system
more two-dimensional, it is believed that increase in $H$
corresponds to an increase in the effective strength of disorder.
A schematic of experimentally observed phase diagram is illustrated in
Fig.\ \ref{fig12}.

\begin{figure}
\caption{The nature of phase transitions at weak ($1^{st}$-order) and strong
($2^{nd}$-order) magnetic fields is illustrated. The diagram is based on
combination of theoretical arguments and recent experiments results (see
text). Tricritical point separating two kinds of transitions at approximately
$10$ Tesla (depending on the strength of disorder) is indicated.}
\label{fig12}
\end{figure}

\subsection{Scaling theory near continuous transition}

The transition from the superconducting Meissner phase to the
non-superconducting
vortex liquid is continuous since the density of vortex
lines (related to the order parameter) increases smoothly at
$H_{c1}$. Also we have argued in the previous sections that at high
applied fields and/or strong disorder the transition from the
superconducting vortex glass phase to the vortex liquid is second
order. As a continuous transition is approached a correlation length
$\xi(T)$, characterizing the length scale over which correlations in
the order parameter exist, begins to grow, diverging exactly at the
transition $T=T_c$ with a critical exponent $\nu$
\beqn\label{corrLength}
\xi(T)={\xi_0} |(T-T_c)/T|^{-\nu}
\eeqn
To the correlation length corresponds a correlation time $\tau$
over which the correlations of size $\xi$ decay.
In the critical region the correlation length is usually
larger than any other physical length in the problem and $\tau$
scales with $\xi$ as $\tau\sim\xi^z$. Based on these
observation a continuous transition can be studied using a scaling
theory, where our ignorance about the transition can be packaged
into a small set of unknown universal critical exponents. Above
we have confined our analysis to an isotropic situation with a single
diverging correlation length. For anisotropic transitions, such as
for example the transition to anisotropic Bose glass (due to correlated
disorder) our analysis can be easily generalized by keeping
track of scaling with correlation lengths in all inequivalent
directions\ \cite{NelsonVinokur92}.

The scaling theory of nonlinear resistivity can be established by
identifying the scaling of the characteristic current $j$ and
the electric field ${\cal E}$ with the correlation length $\xi$, and
constructing a dimensionally-correct combination from these physical
quantities. By definition, in the free-energy a supercurrent couples
to the gauge invariant superfluid velocity and therefore the free-energy
of the superconductor will have an additional current contribution,
\beqn\label{Fcurrent}
\delta F_j=-\int d^d r\ {\phi_0\o
c}\ \vec{j}\cdot\left(\vec{\nabla}\phi-{2\pi\A\o\phi_0}\right)\;.
\eeqn
The long-range correlations in the superconducting order parameter
$\psi\def\|\psi|e^{i\phi}$ are lost beyond the coherence length
scale. Equivalently, the phase $\phi$ changes by $\approx 2\pi$ on
the scale $\xi$ and therefore $|\vec{\nabla}\phi|\sim 2\pi/\xi$. From above
equation (and dimensional analysis) the characteristic value of the
vector potential $\A$ is
\beqn\label{Apotential}
|\A|\sim{\phi_0\o\xi}\;,
\eeqn
which when combined with equation ${\cal\E}=-c^{-1}\pt\A/\pt t$ gives the
scaling
of the characteristic electric field
\beqn\label{Efield}
{\cal E}\sim{\phi_0\o c\xi^{1+z}}\;.
\eeqn
Similarly, from Eq.\ref{Fcurrent} we find that the
characteristic supercurrent density in the correlation volume is
\beqn\label{Jcurrent}
j\sim{c\T\o\phi_0\xi^{d-1}}\;.
\eeqn
In three-dimensions
above equation has a simple interpretation as a requirement
that in the critical region the Magnus energy, in which current $\j$
couples to the projected area $\xi^2$, be of the order $\T$.

Combining above equations we obtain a scaling form of the nonlinear
resistivity for uniform current, valid near the NS transition
\beqn\label{Resistivity}
\rho_{nl}(T,j)={{\cal E}\o j}\approx
\xi^{d-z-2}\tilde{F}_\pm\left({\phi_0\o
c\T}\ j\xi^{d-1}\right)\;,
\eeqn
where the scaling functions $\tilde{F}_{+}$, $\tilde{F}_{-}$
describe the nonlinearity in resistivity for $T>T_c$ and $T<T_c$,
respectively. Above scaling equation can be recast in a more convenient
form in terms of $T$ and $j$ using Eq.\ref{corrLength}
\beqn\label{ResistivityT}
\rho_{nl}(T,j)={{\cal E}\o j}\approx
|T-T_c|^s G_\pm\left({j\o|T-T_c|^\zeta}\right)\;,
\eeqn
where,
\beqna\label{Exponents}
&s&=\nu(2+z-d)\;,\\
&\zeta&=\nu(d-1)\;,
\eeqna
and $G_\pm(x)$ are some other scaling functions.

The frequency dependent linear resistivity is finite
everywhere and the scaling analysis similar to the above leads to
\beqn\label{FrequencyR}
\rho_{l}(T,\omega)\approx\xi^{d-z-2}\tilde{G}_\pm
\left(\omega\xi^z\right)\;.
\eeqn

The properties of the phases above and below $T_c$
gives us the limiting forms of these scaling functions. Above $T_c$
the material is normal and the behavior is Ohmic at small $j$.
Hence,
\beqn\label{OhmicR}
G_{+}(x)\rightarrow {\rm constant},\;\;\; {\rm for}\;\;
x\rightarrow 0\;,
\eeqn
with the dissipation becoming nonlinear for $j\approx
j_{nl}\sim(T-T_c)^\zeta$. The linear resistivity for $j\ll j_{nl}$
is therefore expected to vanish as the transition is approached
\beqn\label{Rvanish}
\rho_l(T)\sim(T-T_c)^s\;.
\eeqn

On the other hand for $j\ll j_{nl}$, i.e. in the limit of
$T\rightarrow T_c$ at $j\neq 0$, we expect nonlinear resistivity to
be finite at $T_c$. This requires
\beqn\label{NonlinearR}
G_\pm(x)\rightarrow x^{s/\zeta}\ ,\;\;\; {\rm
for}\;\;x\rightarrow\infty\;,
\eeqn
so that the divergences in $(T-T_c)$ cancel out. At $T_c$ we find
nonlinear current-voltage relation
\beqn\label{NonlinearIV}
{\cal E}(j)\sim j^{1+s/\zeta}\;,
\eeqn
that is a power-law just like in the Kosterlitz-Thouless
transitions\ \cite{Ambegaokar80}

Finally, below $T_c$ the power law iv-characteristics is replaces by a
stretched exponential, described in Sec.4,
Eq.(\ref{muExp}),
\beqn\label{RbelowTc}
G_{-}(x)\rightarrow e^{-a/x^{\mu}},\;\;\; {\rm for}\;\;x\rightarrow 0\;.
\eeqn

The scaling behavior emerging from above equations have been seen in
{\it dirty}, heavily twinned $Y Ba_2 Cu_3 O_7$
samples\ \cite{Koch89,Gammel91} with an impressive scaling over $4$
decades for the linear resistivity above $T_c$ and over $2$ decades
for $\rho_{nl}(j)$ at the vortex glass-liquid NS transition.
They find $s\approx 6.5$ and $2\nu\approx 4$, critical exponents that are
consistent with the values obtained from numerical
studies\ \cite{Reger91,Gingras92}.

Just as for normal metals skin depth for magnetic fluctuations
(photons) at frequency $\omega$ is
\beqn
\lambda(\omega)\sim\left({|\rho_l(\omega)|\o\omega}\right)^{1/2}\;.
\eeqn
In the superconducting phase $\rho_l(\omega)\sim i\omega/\rho_s$
and therefore
\beqn
\lambda(\omega\rightarrow 0)\sim{1\o\rho_s^{1/2}}
\eeqn
Since near $T_c$
we found that $\rho_l\sim\xi^{1-z}$ and the characteristic frequency
$\omega_\xi\sim\xi^{-z}$, we conclude that $\lambda\sim\xi^{1/2}$ and the
Ginzburg-Landau parameter is
\beqn\label{Kappa}
\kappa={\lambda\o\xi}\sim\xi^{-1/2}\;.
\eeqn
This means that even if the superconductor is characterized by type
II ``bare'' properties, as the transition is approached
and $\xi(T)\rightarrow\infty$, the superconductor becomes
effectively type I. This crossover was first investigated analytically by
Halperin and coworkers\ \cite{Halperin74} who found that the charge
is strongly relevant at the NS transition. Although initial
indications based on renormalization group calculations suggested
that the fluctuations drive the transition to be first-order,
subsequent numerical analysis points to the transition being
that of the inverted XY-model\ \cite{Dasgupta81},
although the story is not fully settled. Recent self-consistent analytical
methods also suggest a continuous second-order transition, but with different
exponents\ \cite{NSradz93}. If the bare $\kappa_{MF}=100$, the crossover to
this new critical behavior will occur very close to the transition, when
$\xi(T)$ has grown by a factor of $10^4$ from what it was when the critical
regime was first entered. Unfortunately, this is experimentally
too close to the transition to be detectable by the presently
available techniques.

\section{CONCLUSION}

As we have seen, the vortex states of high-T$_c$ superconductors are highly
correlated and strongly fluctuating disordered systems that therefore share
much in common with other systems well studied and of much interest in
condensed matter physics.  Many analytical tools have been taken over from
widely different fields of polymer physics, magnetic spin-glass systems,
strongly interacting electrons and bosons in random environment, to name a
few. As we described in these lectures, much progress in understanding
these vortex states has already been accomplished, and a theoretical picture
consistent with experiments and numerical simulations has emerged.
Because the physics of the novel high temperature superconductors
incorporates the most developed and exciting branches of modern condensed
matter physics we expect many further exciting developments and surprises
in our understanding of the phenomenology of high-T$_c$ superconductors.


\begin{thebibliography}{99}
\bibitem{HigherOrder}The higher order terms not included in $F_{GL}$ become
important at low temperatures and in clean samples.
\bibitem{SymmBrake} To be precise, in the ordered superconducting
phase the symmetry is nonlinearly realized or hidden and is not really broken.
\bibitem{commentIsotropic} Although high-T$_c$ materials are highly anisotropic
in that the superconductivity is much stronger in the $ab$-plane than along the
$c$-axis, the transport properties within the $Cu O_2$ (ab) planes is fairly
isotropic.
\bibitem{FluctComment}Runaway renormalization-group flows cannot always be
interpreted as a fluctuation induced first-order transition. For example
normal-superconducting transition in zero magnetic fields
also shows runaway flows near $d=4$ but it is believed to
be second-order of inverted XY universality class in $d=3$ (see
Ref.\ \cite{Halperin74,Dasgupta81}).
\end{thebibliography}
\end{document}